\documentstyle[aps]{revtex}
\input{epsf.tex}
\begin{document}
\draft
\title{Shift of the molecular bound state threshold in dense ultracold 
Fermi gases with Feshbach resonance}
\author{R. Combescot}
\address{Laboratoire de Physique Statistique,
 Ecole Normale Sup\'erieure*,
24 rue Lhomond, 75231 Paris Cedex 05, France}
\date{Received \today}
\maketitle

\begin{abstract}
We consider a dense ultracold Fermi gas in the presence of a Feshbach 
resonance. We investigate how the treshold for bound state formation, 
which is just at the Feshbach resonance for a dilute gas, is modified due 
to the presence of the Fermi sea. We make use of a preceding 
framework of handling this many-body problem. We restrict ourselves 
to the simple case where the chemical potential $ \mu  $ is negative, 
which allows us to cover in particular the classical limit where the effect 
is seen to disappear. We show that, within a simple approach where 
basically only the effect of Pauli exclusion is included, the Fermi sea 
produces a large shift of the threshold, which is of order of the width of 
the Feshbach resonance. This is in agreement with very recent 
experimental findings.
\end{abstract}

After the original breakthrough in the production of Bose-Einstein 
condensates in ultracold bosonic atomic gases and the remarkable 
progress in understanding the underlying physics of these superfluids 
\cite{dgps,cohen}, the exploration of similar dense 
\cite{jin,truscott,schreck} fermionic systems has taken off recently. A 
first series of experiments have reached the strongly degenerate regime 
with mixtures of fermions in two different hyperfine states 
\cite{ketterle,thomas,jin1,bourdel}. One of the major goals in exploring 
these systems is the search for a transition to a BCS superfluid 
\cite{stoofal,rc}, which is the equivalent of Bose condensation for these 
fermionic gases. A major advantage of these gaseous systems is the 
possibility of a direct experimental control of the interaction, through 
electric fields or laser induced resonance for example. However the 
most currently used method is to apply a static magnetic field and go in 
the vicinity of a Feshbach resonance \cite{timm} where the scattering 
length, directly linked to the interaction, is strongly varying.

In this paper we consider how the threshold for formation of the 
molecular bound state linked to this Feshbach resonance is modified by 
the presence of the Fermi sea. The standard Feshbach resonance is due 
to the coupling of an open channel, corresponding to two atoms with 
nuclear and electronic spin degrees of freedom corresponding to a large 
distance situation, with a closed channel where these degrees of 
freedom are in another configuration corresponding to a bound state 
with atoms at short distance, i.e. a molecular state. Actually when the 
energy of this bound state is positive (this corresponds to a positive 
detuning), this bound state is only a quasi-bound state because the 
molecular state can decay into two free atoms, because precisely of the 
small coupling between the two channels. On the other hand when the 
energy of the molecular state is negative (negative detuning), we have a 
true bound state since there is no decay energetically allowed. Naturally 
the existence of this molecular state is important both for fundamental 
and for practical reasons. On the fundamental side it has been 
emphasized recently \cite{holland} that the strong interaction produced 
by the vicinity of this resonance could be a major help for the BCS 
transition. Moreover the physics would become richer, due to the 
appearance of a Bose condensate of the molecules corresponding to the 
bound state. On the practical side the production of this molecular state 
is a first step in the decay of the whole metastable atomic gas, produced 
by three-body recombination \cite{3body}. Therefore it is of clear 
interest to investigate what happens to this molecular state due to the 
influence of the Fermi sea.

Quite recently  \cite{rcfesh}  we have considered how the scattering 
amplitude due to a Feshbach resonance is modified by the presence of 
the dense Fermi gas. Note that the problem of the modification of 
scattering by interactions in cold quantum gases has already been 
adressed by H. Stoof \cite{stoof} .
Although the formalism we used \cite{rcfesh} is quite general, 
its application in this paper has been strongly simplified. First the 
irreducible vertex has been taken merely as the simple scattering 
amplitude of two isolated atoms. Second the effect of the Fermi gas has 
been taken into account by including only  Pauli exclusion. Even so the 
resulting scattering amplitude displays quite non trivial features. First, 
in contrast with what happens in vacuum, it depends on the total 
momentum of the scattering atoms. This is easily understood since, for 
two atoms with very high momentum, the blocking of some states due 
to Pauli exclusion will only alter slightly their scattering. On the other 
hand when the total momentum is zero and the energy of the atoms is of 
the order of a typical energy for atoms in the Fermi gas, many states 
which are relevant for the scattering of the two atoms in vacuum will be 
blocked due to Pauli exclusion and the scattering amplitude is deeply 
modified. In the following we consider only this case of zero total 
momentum since it displays the strongest manifestation of Pauli 
exclusion. Also, just as in \cite{rcfesh}, we omit for simplicity any 
background scattering, so the scattering originates only from the 
Feshbach resonance.

In this case the following result \cite{rcfesh}  is obtained for the inverse 
of the scattering amplitude $f( \omega )$ as a function of the total 
energy $ \omega $ of the scattering particles (we take $ \hbar= 1$ and $ 
k _{B}=1$):
\begin{eqnarray}
\frac{1}{ f ( \omega)} = - \frac{1}{a} - \frac{\omega }{ \gamma } +  
\frac{2}{ \pi } \int_{0}^{ \infty}dk [  1 - \frac{k ^{2}}{ k ^{2} - m 
\omega }  \tanh  \frac{ \epsilon _{k}- \mu }{2 T}]  
\label{eq1}
\end{eqnarray}
with $ \epsilon _{k} = k ^{2}/2m$. Here the coupling parameter $ 
\gamma $ is also a measure of the energetic width of the Feshbach 
resonance, while $ a$ is the scattering length due to the Feshbach 
resonance which is directly linked to the detuning $ \omega _0$ by $ a 
= - \gamma / \omega _0$. In contrast with Ref. \cite{rcfesh}, the origin 
for the energy $ \omega $ of the particles has not been shifted at the 
chemical potential. This energy $ \omega $ has to be understood with 
an infinitesimal positive imaginary part. One recovers the case of two 
atoms in vacuum by setting the chemical potential $ \mu =0$ and 
looking at the $ T = 0$ limit. Then the last term gives only the 
contribution required by unitarity and one finds as expected:
\begin{eqnarray}
f ( \omega) = - \frac{1}{(\omega - \omega _0 ) / \gamma + i (m \omega 
) ^{1/2}}
\label{eq2}
\end{eqnarray}
The standard Feshbach resonance, with a diverging scattering 
amplitude, is found at zero energy $ \omega = 0$ when the detuning $ 
\omega _0$ is equal to zero. This corresponds to an infinite scattering 
length $a$. In contrast nothing peculiar happens in Eq.(1) when one 
crosses the infinite scattering  length $ a ^{-1}=0$ situation. Actually in 
the case of $^6$Li the coupling $ \gamma $ is very large and we are left 
only with the last term in Eq.(1) which originates from the presence of 
the Fermi sea. Clearly when $ a ^{-1}$ becomes small, it becomes also 
irrelevant. This is in agreement with the recent experimental results of 
various groups \cite{ketterle,thomas,jin1,bourdel} who do not see any 
accident when they cross the supposed location of the resonance. This 
finding is also completely coherent with a very recent calculation of 
Pitaevskii and Stringari \cite{pitstring} where they showed that the 
virial coefficient has no discontinuity when one goes across the unitary 
limit $ a ^{-1}=0$.

In this paper we want to consider another immediate consequence of 
Eq.(1) which is also directly relevant for experiments. For the case 
Eq.(2) of two particles in vacuum the appearance of the bound state of 
these two particles, corresponding physically to the existence of a 
molecular state, occurs as soon as the detuning becomes negative, that 
is when the scattering length becomes positive $ a > 0$. Experimentally 
this corresponds to a well defined magnetic field, since the detuning is 
directly controlled by the applied magnetic field. This molecular state 
appears as a pole in $f( \omega )$ as given by Eq.(2) for negative 
values of the energy. One finds specifically for the binding energy $ 
\epsilon _b = - \omega $ of this state:
\begin{eqnarray}
\epsilon _b = \frac{1}{m a ^{2}} \frac{2}{1+2r+\sqrt{1+4r}} 
\label{eq3}
\end{eqnarray}
where $ r = (1/ma^2)/ | \omega _0| = 1/(m \gamma a)$ is the ratio 
between the two limiting values of this binding energy. Indeed for $ r \ll 
1$ one finds the standard binding energy $ \epsilon _b = 1/ma^2$ 
associated with the positive scattering length $a$, while for $ r \gg 1$ 
the binding energy is equal to the detuning $ \epsilon _b = | \omega 
_0|$.

Now it is clear that the existence of the Fermi sea will shift in general 
the magnetic field at which the molecular state appears. This is 
specifically produced by the last term in Eq.(1). Indeed it is seen that 
for $ \omega = 0$ one does not find a pole for $ a ^{-1}=0$ since in 
this case the last term is always positive because $ \tanh x < 1 $. Here 
we restrict ourselves to the simple case where the chemical potential $ 
\mu  $ is negative, which allows us to cover in particular the classical 
limit where the effect is seen to disappear. In order to find the magnetic 
field $B$ for which the molecules appear we look for the corresponding 
value of the scattering length $ a(B)$. We find it from Eq.(1) by 
assuming that the binding energy of the molecules is zero when they 
first appear. We will afterwards check this hypothesis. Since we are 
interested in non-positive values of the energy, the imaginary part in 
Eq.(1) is zero and we have only to deal with the real part. Setting Re$ f 
^{-1}( \omega) = 0$ for $ \omega =0$ leads to:
\begin{eqnarray}
\frac{1}{a} =  \frac{4}{ \pi } \int_{0}^{ \infty}dk \frac{1}{e ^{\frac{ 
\epsilon _{k}+ | \mu | }{T}} + 1}
\label{eq4}
\end{eqnarray}
At the level of our approximation the chemical potential $ \mu $ in 
Eq.(3) is merely linked to the single species particle density $ n$ by the 
free particle equation:
\begin{eqnarray}
n = \frac{1}{2 \pi ^2} \int_{0}^{ \infty}dk \frac{ k ^{2}}{e ^{\frac{ 
\epsilon _{k}+ | \mu | }{T}} + 1}
\label{eq5}
\end{eqnarray}
In the classical limit $ e ^{| \mu | /T} \rightarrow \infty $ one finds 
easily from Eq.(4):
\begin{eqnarray}
\frac{1}{a} = 4 ( \frac{mT}{2 \pi })^{1/2} e ^{-\frac{ | \mu | }{T}} =  
\frac{4}{ \Lambda _T } e ^{-\frac{ | \mu | }{T}}
\label{eq6}
\end{eqnarray}
where $ \Lambda _T  =  (2 \pi \hbar ^{2}/mT)^{1/2} $ is the de 
Broglie thermal wavelength, while Eq.(5) gives the well known 
classical ideal gas relation:
\begin{eqnarray}
n =  ( \frac{mT}{2 \pi })^{3/2} e ^{-\frac{ | \mu | }{T}} =  \Lambda 
_T ^{-3} e ^{-\frac{ | \mu | }{T}}
\label{eq7}
\end{eqnarray}
Together these results lead to:
\begin{eqnarray}
\frac{1}{a} = 8 \pi  \frac{n \hbar ^{2}}{mT} = 4 n \Lambda _T ^{2}
\label{eq8}
\end{eqnarray}
where we have reintroduced the Planck constant. As expected $ a ^{-
1}$ goes to zero in the limit of large temperature and low density, and 
we recover that the molecular state appears right below the Feshbach 
resonance. We can also rewrite this result by introducing a wavevector 
$ k_F$ linked to the density $n$ by the same relation as in the $T=0$ 
limit, namely $ n = k_F ^{3}/6 \pi ^2$. If we set $ E_F = \hbar ^{2} 
k_F ^{2}/2m$ for the corresponding energy we have:
\begin{eqnarray}
k_F a = \frac{3 \pi }{8} \frac{T}{E_F}
\label{eq9}
\end{eqnarray}
Naturally the classical regime corresponds to $ T \gg E_F $, so the 
scattering length $a$ at the threshold for molecule formation gets very 
large, all the more since, for a trapped gas, the density $n$ decreases 
when the temperature increases which makes the length scale $ 1/k_F$ 
(basically the interparticle distance) increase with $T$. On the other 
hand the extrapolation of this simple formula in the regime $ T \sim 
E_F$ shows immediately that the molecular threshold is shifted in this 
case to a scattering length $ a \sim  1/k_F $. In the specific case of 
$^6$Li , one of the most studied experimentally, this gives for the 
degenerate conditions a scattering length of order of 100 nm. This 
implies a shift corresponding to a change of the magnetic field of order 
of the width of the Feshbach resonance, which is typically 100G. This 
large shift is in very good qualitative agreement, both in sign and in 
magnitude, with the very recent experimental results of Dieckmann et 
al. and of Bourdel et al. \cite{bourdel}.

\begin{figure}
\centering
\vbox to 65mm{\hspace{-6mm} \epsfysize=65mm 
\epsfbox{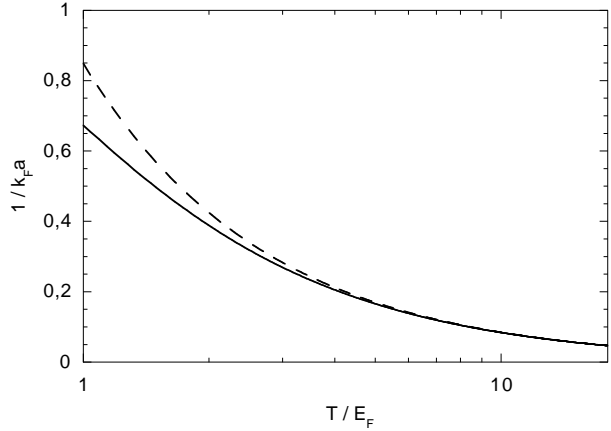} }
\caption{Reduced inverse scattering length $ 1/ (k_F a) = (6 \pi  ^{2} n 
a ^{3}) ^{-1/3}$ at the bound state threshold as a function of the 
degeneracy parameter $R=T/E_F=(16/9 \pi ) ^{1/3}/(n \Lambda _T  
^{3}) ^{2/3}$, with a logarithmic scale for $R$. Full line: numerical 
result Eq.(10-11). Dashed line: asymptotic result Eq.(9).}
\label{figure1}
\end{figure}

In order to be more quantitative in the general case we have just to 
handle Eq.(4) and (5) numerically. This is done conveniently by taking 
$y = k / (2mT) ^{1/2}$ as new variable. Introducing $k_F$ as above, 
setting $ R = T / E_F$ and $ A = \exp ( | \mu | /T)$, we see that Eq.(5) 
gives $R$ in terms of $A$ as:
\begin{eqnarray}
R ^{-3/2} = 3 \int_{0}^{ \infty}dy \frac{ y ^{2}}{ A e ^{y ^{2}} + 
1}
\label{eq10}
\end{eqnarray}
while Eq.(4) then leads for the threshold scattering length $a$ to:
\begin{eqnarray}
\frac{1}{k_F a} =  \frac{4}{ \pi } R ^{1/2} \int_{0}^{ \infty}dy 
\frac{1}{ A e ^{y ^{2}} + 1}
\label{eq11}
\end{eqnarray}
For our range of interest $A \ge 1$, our degeneracy parameter $R$ is 
going from $\sim 1$ to $ \infty$. The result for $ 1/k_F a $ as a 
function of $ R = T / E_F$ is given in Fig. 1 with a logarithmic scale 
for $R$. We have also plotted the asymptotic result Eq.(9) which is 
seen to give still a reasonable estimate for $R=1$.

We have still to check that the threshold corresponds to molecules with 
zero binding energy. If we look for a bound state with binding energy $ 
\epsilon _b$ appearing for a scattering length $a_b$, we have from 
Eq.(1) with $ \omega = - \epsilon _b$:
\begin{eqnarray}
\frac{1}{a_b} =  \frac{\epsilon _b }{ \gamma } +  \frac{2}{ \pi } 
\int_{0}^{ \infty}dk [  1 - \frac{k ^{2}}{ k ^{2} + m \epsilon _b }  
\tanh  \frac{ \epsilon _{k}+  | \mu | }{2 T}]  
\label{eq12}
\end{eqnarray}
Taking the difference with Eq.(4) which gives the scattering length $a$ 
for zero binding energy, we find:
\begin{eqnarray}
\frac{1}{a_b} - \frac{1}{a}=  \frac{\epsilon _b }{ \gamma } +  
\frac{2}{ \pi } \int_{0}^{ \infty}dk \frac{ m \epsilon _b }{ k ^{2} + m 
\epsilon _b }  \tanh  \frac{ \epsilon _{k}+ | \mu | }{2 T} 
\label{eq13}
\end{eqnarray}
which is positive. So, just as for the case of two particles in vacuum, 
molecules with non zero binding energy correspond to lower scattering 
length than those with zero binding energy. Therefore they appear later 
when one starts from the Feshbach resonance and decreases 
progressively the scattering length by lowering the magnetic field.

Naturally we have to keep in mind that this result represents the 
maximum shift on the threshold for the molecular state, since it 
corresponds to the case where the total momentum is zero. There will 
be a similar shift for non zero momentum, but it decreases with 
momentum as we have already mentionned since for very high 
momentum the blocking effect of the other fermions is clearly very 
weak. So there is actually a distribution of threshold with $a ^{-1}$ 
going from zero to its maximum value given by Eq.(4). Similarly we 
have in real experimental systems a distribution of densities due to the 
trapping potential. So direct comparison with experiment should take 
into account these two distributions. Nevertheless the maximum value 
of the threshold is given by Eq.(4) with the particle density equal to its 
maximum value at the center of the trap. Starting from low positive 
values of the scattering length and increasing it, which corresponds for 
$^6$Li to increase the magnetic field starting from low values, our 
result Eq.(4) corresponds therefore to the lowest magnetic field for 
which resonating conditions for the formation of molecules will occur. 
Of course molecular bound state do exist for all the range $a>0$ but it is 
known that its formation through three-body recombination 
\cite{3body} is strongly enhanced in the immediate vicinity of the 
resonant situation. Note that important dynamical effects will necessarily
occur in experiments, which will make quantitative comparison with
theory even harder. In this respect it seems that direct spectroscopic
observation of the molecules would provide the best investigation tool.

In conclusion we have studied the effect of Pauli exclusion due to the 
Fermi sea on the Feshbach resonance. We have shown that, as a natural 
consequence, there is a shift of the threshold for the formation of 
molecules under resonating conditions. The sign and the size of the 
effect is in very good qualitative agreement with the very recent 
experimental findings of Dieckmann et al. and of Bourdel et al. 
\cite{ketterle,bourdel}. Naturally it must be kept in mind that our 
theoretical treatment is rough since we take only into account Pauli 
exclusion and we have not included the effect of the interactions in the 
last term in Eq.(1). Nevertheless we do not expect the physics linked to 
Pauli exclusion to disappear in the full solution of the problem. Proper 
account of the interactions is likely to modify quantitatively the size of 
the shift, but we do not expect them to modify it qualitatively. In 
particular it is unlikely to suppress this effect and make this shift almost 
disappear. Therefore we believe that our simple estimates provides 
already a proper physical understanding of the shift.

We are very grateful to T. Bourdel, Y. Castin, C. Cohen-Tannoudji, J. 
Dalibard, X. Leyronas, C. Mora, C. Salomon and G. Shlyapnikov for 
very stimulating discussions.

* Laboratoire associ\'e au Centre National
de la Recherche Scientifique et aux Universit\'es Paris 6 et Paris 7.

\end{document}